\documentclass[11pt,epsbox]{article}
\usepackage[dvips]{graphicx}
\usepackage{amsmath,amssymb}

\baselineskip=\normalbaselineskip
\renewcommand{\baselinestretch}{1.4}
\makeatletter
\@addtoreset{equation}{section}

\makeatother
\newcommand{\be}{\begin{equation}}
\newcommand{\ee}{\end{equation}}
\newcommand{\ba}{\begin{eqnarray}}
\newcommand{\ea}{\end{eqnarray}}
\newcommand{\nn}{\nonumber \\}

\newcommand{\del}{\partial}
\newcommand{\bra}[1]{\left\langle\,{#1}\,\right|}
\newcommand{\ket}[1]{\left|\,{#1}\,\right\rangle}

\newcommand{\tb}{\widetilde{b}}

\newcommand{\talpha}{\widetilde{\alpha}}

\newcommand{\bN}{\bar{N}}

\newcommand{\bX}{{\mathbf X}}

\newcommand{\bP}{\mathbf P}
\newcommand{\bM}{\mathbf M}

\newcommand{\bGamma}{\mathbf \Gamma}

\newcommand{\hx}{\widehat{x}}

\newcommand{\hp}{\widehat{p}}

\newcommand{\hsigma}{\widehat{\sigma}}

\newcommand{\bx}{\mathbf x}
\newcommand{\bp}{\mathbf p}

\newcommand{\Tr}{{\rm Tr}}

\begin{document}
\setcounter{page}{0}
%%%%%%%%%%%%%%%%%%%

\begin{flushright}
\parbox{40mm}{%
RESCEU-11/05 \\
RIKEN-TH-47 \\
hep-th/0506221 \\
June 2005}
\end{flushright}

\vfill

%%%%%%%%%%%%%%%%%%%%%%%%%%%%%%%%%%%%%%%%%%%%%%%%%%%%
%% Title
%%%%%%%%%%%%%%%%%%%%%%%%%%%%%%%%%%%%%%%%%%%%%%%%%%%%
\begin{center}
{\Large{\bf 
Excited D-branes and Supergravity Solutions
}}
\end{center}

\vfill

\renewcommand{\baselinestretch}{1.0}

%%%%%%%%%%%%%%%%%%%%%%%%%%%%%%
%% author
%%%%%%%%%%%%%%%%%%%%%%%%%%%%%%
\begin{center}
\textsc{Tsuguhiko Asakawa}
\footnote{E-mail: \texttt{t.asakawa@riken.jp}}, 
\textsc{Shinpei Kobayashi}
\footnote{E-mail: \texttt{shinpei@resceu.s.u-tokyo.ac.jp}} and
\textsc{So Matsuura}
\footnote{E-mail: \texttt{matsuso@riken.jp}} 
%\\[2em]

~\\
$^{1, 3}$ \textsl{Theoretical Physics Laboratory, \\ 
      The Institute of Physical and Chemical Research (RIKEN), \\
     2-1 Hirosawa, Wako, Saitama 351-0198, JAPAN } \\ 

\vspace{0.5cm}
$^2$ \textsl{Research Center for the Early Universe, \\ 
      The University of Tokyo,  
     7-3-1 Hongo, Bunkyoku, Tokyo 113-0033, JAPAN}  
\end{center}

%%%%%%%%%%%%%%%%%%%%
%% abstract
%%%%%%%%%%%%%%%%%%%%
\begin{center}
{\bf abstract}
\end{center}

\begin{quote}

\small{%
We investigate the general solution 
with the symmetry $ISO(1,p) \times SO(9-p)$  of Type II supergravity 
(the three-parameter solution) from the viewpoint of the superstring
theory. 
We find that one of the three parameters ($c_1$) 
is closely related to the ``dilaton charge'' 
and the appearance of the dilaton charge is a consequence
of deformations of the boundary condition from that of the boundary 
state for BPS D-branes.
We give three examples of the deformed D-branes by considering 
the tachyon condensation from systems of D$p$\=D$p$-branes, 
unstable D9-branes and unstable D-instantons 
to the BPS saturated D$p$-branes, respectively. 
We argue that the deformed systems are generally regarded as 
tachyonic and/or massive excitations of the open strings
on D$p$\={D}$p$-brane systems.
}
\end{quote}
\vfill

\renewcommand{\baselinestretch}{1.4}

%%%%%%%%%%%%%%%%%%%%%%%%%%%%%%%%%%
% Main
%%%%%%%%%%%%%%%%%%%%%%%%%%%%%%%%%%
\renewcommand{\thefootnote}{\arabic{footnote}}
\setcounter{footnote}{0}
\addtocounter{page}{1}
%%%%%%%%%%%%%%%%%%%%%%%%%%%%%%%%%
\newpage
\section{Introduction and Overview}
\label{sec:intro}

Black holes are important objects to investigate the nature
of space-time beyond the description by the general relativity. 
For example, the study of the black hole entropy provides 
the holographic principle \cite{Susskind:1994vu}, 
which gives a strong constraint to the property of quantum gravity. 
The microscopic properties of black holes are also 
intensively investigated 
from the viewpoint of string theory. 
For example, many black hole (or brane) solutions are known 
in the low energy effective theory of the superstring theory. 
In particular, for BPS saturated solutions of Type II 
(or Type I) supergravity, 
it is well known that there is a correspondence between the classical
solutions and the systems of the BPS D-branes \cite{Horowitz:1991cd}.  
In this sense, it is natural to assume that there is a stringy origin 
for a general classical solution of Type II supergravity, even for 
non-BPS systems. 

In our previous paper \cite{Kobayashi:2004ay}, 
we treated a general classical solution 
of Type II supergravity with the symmetry 
$ISO(1,p) \times SO(9-p)$, 
which is called as the ``three-parameter solution'' 
\cite{Zhou:1999nm}. 
The solution had been thought to be 
the low energy counterpart of the D$p$\={D}$p$-brane system 
with a constant tachyon profile (vacuum expectation value (VEV)), 
where the three parameters had been supposed to 
correspond to the number of the D-branes, that of the anti D-branes
and the VEV of the tachyon field 
\cite{Brax:2000cf, Lu:2004ms}. 
In \cite{Kobayashi:2004ay}, however,  
we re-examined this correspondence and found that the above conjecture is
not correct and the three-parameter solution is the low energy counterpart
of the D$p$\={D}$p$-brane system with a constant tachyon VEV  
only if one of the three parameters, $c_1$, is tuned to zero. 

In this paper, we consider the stringy interpretation 
of missing $c_1$ and 
give some explicit examples of the stringy counterparts of 
the three-parameter solution.
As we mentioned in the previous paper
\cite{Kobayashi:2004ay}, $c_1$ is related to the ``dilaton charge'', 
which characterizes the asymptotic behavior of the dilaton field. 
It is well known that black holes with horizons do not have 
the dilaton charge in general, which is supported by 
the no-hair theorem, but some exceptions are known 
\cite{Horowitz:1991cd,Gibbons:1987ps,Garfinkle:1990qj} and 
we have not understood this phenomena 
systematically \cite{Agnese:1985xj}. 
One of the purposes of this paper is to give an approach to this problem
from the microscopic (string theoretical) point of view. 

Our strategy is making use of the known correspondence between 
the BPS black $p$-brane solution in the supergravity 
and the BPS D$p$-branes in the string theory,
and extending this correspondence by appropriately deforming 
the both sides. 
On the supergravity side, this leads to a new characterization of 
the solution.
We will find that the BPS limit of the three-parameter solution
consists of two extremal limit, that is,
the ordinary BPS condition and the condition that the dilaton charge 
is proportional to the ADM mass.
Then, the three-parameter solution is characterized by two non-extremality 
parameters (and the RR-charge).
In particular, the parameter $c_1$ is related to the second non-extremality.
On the string theory side, we will formally deform the 
boundary state for the BPS D$p$-brane, which respects the same symmetry
of the three-parameter solution. 
By evaluating the long distance behavior of the solution of the 
supergravity and the massless emission from the deformed boundary state, 
we will directly compare the (microscopic) deformation parameters 
of the boundary state 
and the (macroscopic) parameters in the three-parameter solution. 
From this analysis, we show that 
the first kind of non-extremality corresponds to the non-BPS nature,
while the second kind (hence $c_1$) is related to the deformation of 
the boundary condition from the ordinary Neumann/Dirichlet one.

As examples of the deformed boundary states, we consider three systems of
D-branes. 
The first one is the system we consider in \cite{Kobayashi:2004ay},
which possesses only the first non-extremality since 
the D$p$\={D}$p$-system follows the ordinary boundary
condition, that is, the Neumann boundary condition for the longitudinal
directions and the Dirichlet boundary condition for the transverse
directions. 
The second and the third example are systems of unstable D9-branes and
unstable D-instantons, respectively, 
with an appropriate tachyon condensation to the BPS D$p$-branes,
which actually have non-trivial dilaton charges, i.e. non-zero value of $c_1$. 
They possess the deformation of the boundary
condition for the worldsheet, 
which 
represent in some sense fuzziness in transverse or longitudinal directions, 
respectively. 
From these examples, we discuss that 
deformed boundary states which respect the symmetry of the
three-parameter solution can be obtained by considering 
D$p$\={D}$p$-systems on which 
tachyonic and/or massive excitations are turned, in general.

The organization of this paper is as follows. In the next section, 
we review the three-parameter solution and we 
propose a new characterization of the solution in terms of 
two non-extremality parameters.  
In the section 3, 
we give the string theoretical interpretation to the parameter $c_1$ 
and give three examples of the deformed boundary states. 
The section 4 is devoted to the conclusions and discussions.

%%%%%%%%%%%%%%%%%%%%%%%%%%%%%%%%%
\section{The Three-Parameter Solution and Non-Extremalities}
\label{sec:3para}

In this section, we first give a short review of the three-parameter solution.
Then, by investigating the BPS limit of the solution, we 
propose a new characterization of the solution in terms of 
two non-extremality parameters.  

First of all, we consider Type II supergravity 
in the following setting:
\begin{itemize}
\item Assume a fixed, $(p+1)$-dimensional object as a source 
carrying only RR $(p+1)$-form charge.
\item Spacetime has the symmetry $ISO(1,p) \times SO(9-p)$ 
and it is asymptotically flat.
\end{itemize}
Note that these ansatzes are the same as those 
for the BPS black $p$-brane solution 
(we will consider the region $0\le p \le 6$ in this paper).
Under these conditions, it is sufficient to consider 
the metric, the dilaton and the RR $(p+1)$-form field. 
The relevant part of the ten-dimensional 
action (in the Einstein frame) is given by
\begin{equation}
S = \frac{1}{2\kappa^2} \int d^{10} x \sqrt{-g} 
    \left[ 
    R -\frac{1}{2}(\del \phi)^2 
    -\frac{1}{2(p+2)!} e^{\frac{3-p}{2} \phi} |F_{p+2}|^2 
    \right], 
\label{ten-3para-action}
\end{equation}
where $F_{(p+2)}$ denotes the $(p+2)$-form field strength which 
is related to the $(p+1)$-form potential of the RR-field $\mathcal{A}^{(p+1)}$ 
as $F_{(p+2)} = d \mathcal{A}_{(p+1)}$. 
According to the symmetry $ISO(1,p)\times SO(9-p)$, 
we should impose the following ansatz, 
\begin{align}
ds^2 &= g_{MN} dx^M dx^N \nonumber \\ 
     &= e^{2A(r)} \eta_{\mu\nu}dx^{\mu}dx^{\nu} 
      + e^{2B(r)} \delta_{ij} dx^i dx^j \nonumber \\
     &=e^{2A(r)} \eta_{\mu\nu}dx^{\mu}dx^{\nu} 
      + e^{2B(r)} (dr^2 + r^2 d\Omega_{(8-p)}^2 ),  \nonumber \\
\phi &= \phi(r), \nonumber \\ 
\mathcal{A}^{(p+1)} &= e^{\Lambda(r)}\ 
             dx^0 \wedge dx^1 \wedge \cdots \wedge dx^p,  
\label{ten-3para-ansatz}
\end{align}
where $M=(\mu,i)$, $\mu,\nu=0,\cdots,p$ are indices of the longitudinal  
directions of the $p$-brane, 
$i,j=p+1,\cdots,9$ express the orthogonal directions, 
and $r^2=x^i x_i$. 

The authors of \cite{Zhou:1999nm} gave a general solution of Type II supergravity 
of this system.
As the solution includes three integration constants, 
it is called as the three-parameter solution, which is given by%
\footnote{
In \cite{Zhou:1999nm}, the authors consider the $D$-dimensional gravity
theory and have constructed a general solution 
with the less symmetry $SO(p) \times SO(D-p-1)$ 
as the four-parameter solution. 
The three-parameter solution is a restricted one.}  
\begin{align}
\label{sln-1}
A(r) &= \frac{(7-p)(3-p)c_1}{64} h(r) 
        -\frac{7-p}{16} 
        \ln \left[
            \cosh (k h(r)) -c_2 \sinh (k h(r))
            \right], \\
\label{sln-2}
B(r) &= \frac{1}{7-p} \ln [f_-(r) f_+(r)] \nonumber \\
        & \hspace{0.5cm}-\frac{(p+1)(3-p) c_1}{64} h(r) 
        +\frac{p+1}{16} 
        \ln \left[
            \cosh (k h(r)) -c_2 \sinh (k h(r))
            \right], \\
\label{sln-3}
\phi(r) &= \frac{(p+1)(7-p) c_1}{16} h(r) 
           +\frac{3-p}{4} 
            \ln \left[
            \cosh (k h(r)) -c_2 \sinh (k h(r))
            \right], \\
\label{sln-4}
e^{\Lambda(r)} &= -\eta (c_2^2 -1)^{1/2}\  
                  \frac{\sinh(k h(r))}
                   {\cosh (k h(r)) -c_2 \sinh (k h(r))},
\end{align}
%-----------------------------------------%
where 
%-----------------------------------------%
\begin{align}
f_{\pm}(r) &\equiv 1 \pm \frac{r_0^{7-p}}{r^{7-p}}, \\
h(r) &\equiv \ln \left( \frac{f_-}{f_+} \right), \\
k &\equiv \pm
\sqrt{ \frac{2(8-p)}{7-p} - \frac{(p+1)(7-p)}{16} c_1^2 } 
 \nonumber \\
&\equiv \pm \frac{\sqrt{(p+1)(7-p)}}{4}\, \sqrt{c_m^2-c_1^2}, 
\qquad \left(c_m = \sqrt{\frac{32(8-p)}{(p+1)(7-p)^2} } \right), \\
\eta &\equiv \pm 1,  
\end{align}
%-----------------------------------------% 
where $\eta$ denotes the sign of the RR-charge. 
The three parameters% 
\footnote{
We have labeled $c_3$ of \cite{Zhou:1999nm} as $c_2$ and $k$ as $-k$ 
according to \cite{Brax:2000cf}.}, 
$r_0,\ c_1,\ c_2$, are the integration constants 
that parametrize the solution. 
As in \cite{Brax:2000cf}, the domain of the parameters 
in the solution (\ref{sln-1})--(\ref{sln-4}) we take is 
\begin{align}
 c_1 &\in [0, c_m], \\
 c_2 &\in (-\infty,-1]\cup[1,\infty), \\
 r_0^{7-p} &\in {\mathbf R}.
\end{align}
where we have already fixed the ${\mathbf Z}_2$ symmetries 
of the solution, 
\begin{align}
 (r_0^{7-p},c_1,c_2,sgn(k),\eta) &\to 
 (r_0^{7-p},c_1,-c_2,-sgn(k),-\eta),  \nonumber \\
 (r_0^{7-p},c_1,c_2,sgn(k),\eta) &\to 
 (-r_0^{7-p},-c_1,c_2,-sgn(k),\eta), 
\end{align}
by choosing $c_1 \ge 0$ and $k\ge 0$.
Furthermore, we have a degree of freedom to choose 
the signs of $r_0^{7-p}$ and $c_2$. 
As we mentioned in Sec.1, 
we will discuss the BPS black $p$-brane limit 
of the three-parameter solution and in order to take that limit consistently, 
we must choose one of the two branches, 
\begin{eqnarray}
\begin{cases}
r_0^{7-p} \ge 0,\ c_2 > 0 &\mbox{: branch} \ I_+, \\ 
r_0^{7-p} \le 0,\ c_2 < 0 &\mbox{: branch} \ I_-.
\end{cases}
\end{eqnarray} 

{}From the viewpoint of the gravity theory, 
the three-parameter solution describes 
charged dilatonic black objects.  
Thus, the RR-charge $Q$, the ADM mass $M$ and 
the dilaton charge $D$
are natural quantities to characterize the solution. 
For convenience, we consider wrapping the spatial worldvolume 
directions on a torus $T^p$ of volume $V_p$. 
The RR-charge is given by an appropriate surface integral 
over the sphere-at-infinity in the transverse directions 
\cite{Brax:2000cf,Stelle:1998xg};  
\begin{equation}
|Q| =  2 (c_2^2- 1)^{1/2} k N_p r_0^{7-p}, 
\label{RR-charge}
\end{equation} 
where 
\begin{equation}
N_p \equiv \frac{(8-p)(7-p) \omega_{8-p}V_p}{16\kappa^2},  
\label{N_p}
\end{equation}
and $\omega_d=\frac{2\pi^{(d+1)/2}}{\Gamma((d+1)/2)}$ 
is the volume 
of the unit sphere $S^d$. 
The ADM mass is defined in 
\cite{Maldacena:1996ky,Myers:1986un} as
\begin{equation}
g_{00} = -1 + \frac{7-p}{8}\frac{M}{N_p}\frac{1}{r^{7-p}}
            +\mathcal{O}\left( \frac{1}{r^{2(7-p)}}\right) ,
\end{equation}
where (\ref{N_p}) is used.
This gives us \cite{Brax:2000cf} 
%--------------------------------------%
\begin{equation}
M = \left(2c_2 k + \frac{3-p}{2}c_1 \right) N_p r_0^{7-p}, 
\label{ADM-mass}
\end{equation}
Similarly, we define the dilaton charge by its asymptotic behavior 
in the transverse direction as \cite{Ohta:2002ac}
\footnote{
Their original definition is in the string frame but here we
define it in the Einstein frame in the same manner.
} 
\begin{equation}
\phi =  \frac{D}{N_p}\frac{1}{r^{7-p}}
           +\mathcal{O}\left( \frac{1}{r^{2(7-p)}}\right), 
\end{equation}
which gives 
\begin{equation}
D = \left(\frac{3-p}{2}c_2 k -\frac{(p+1)(7-p)}{8}c_1 \right) N_p r_0^{7-p}.
\label{dilaton-charge}
\end{equation}
%--------------------------------------% 

Here we give some historical remarks on the dilaton charge. 
As we referred in \cite{Kobayashi:2004ay}, the dilaton charge 
is a familiar   
quantity in the general relativity \cite{Buchdahl:1959nk,
Janis:1968,Wyman:1981bd,Virbhadra:1997ie} 
and it has been investigated for a long time in various contexts, 
for example, the no-hair theorem \cite{Bekenstein:1975ts}, 
the cosmic censorship,  
the scalar field collapse \cite{Husain:1994uj}. 
It is well known that 
the existence of the dilaton charge would change the structures of 
spacetimes drastically \cite{Wyman:1981bd,Agnese:1985xj,Koikawa:1986qu} 
and generally cause naked singularities to appear% 
\footnote{There are some exceptions which have horizons even if 
there is non-zero dilaton charge 
\cite{Gibbons:1987ps,Garfinkle:1990qj,Horowitz:1991cd} as we mentioned 
in the introduction. 
These solutions may be related not to the three-parameter solution, but to 
the four-parameter solution. The three-parameter  
and the four-parameter solution coincide with each other 
in the case of $D=4$ and $p=0$, so they are not distinguishable 
in such a case. 
We will discuss the relation between the condition of the formation 
of horizons and the parameters of the supergravity solutions in the 
forthcoming paper.}. 
The extensions to the higher-dimensional theories \cite{Xanthopoulos:1989kb} 
or the stabilities of such singular spacetimes have been also investigated 
by many authors \cite{Clayton:1998ay,Clayton:1998zz}. 
In spite of those investigations, we have not systematically understood 
the relation between the horizon and the dilaton charge, for example, 
what kind of the dilaton couplings to the RR-charge can avoid to make 
a spacetime singular.  
%the physical meaning of the dilaton charge 
%is still an open question  
%as pointed out in \cite{Agnese:1985xj}. 

Since the three-parameter solution is the most general solution with the
symmetry $ISO(1,p)\times SO(9-p)$,
it naturally includes the BPS black $p$-brane solution; 
\begin{eqnarray}
ds^2 &=& f(r)^{-\frac{7-p}{8}} \eta_{\mu\nu} dx^{\mu}dx^{\nu} 
       +f(r)^{\frac{p+1}{8}} \delta_{ij} dx^{i}dx^{j}, 
\label{extreme-1}
\\
e^{\phi}(r) &=& f(r)^{\frac{3-p}{4}}, 
\label{extreme-2}
\\
e^{\Lambda(r)} &=& -\eta\left(f(r)^{-1} -1\right),
\label{extreme-3}
\end{eqnarray}
where 
\begin{equation}
f(r) = 1 +\frac{\mu_0}{r^{7-p}}.
\end{equation}
%--------------------------------------------------------%
Note that $\mu_0$ is the only parameter of the solution, 
and therefore, all the quantities are labeled by it as
\begin{equation}
|Q|=M=N_p \mu_0 , \qquad  D=\frac{3-p}{4}N_p \mu_0, 
\end{equation}
where $\mu_0$ takes an arbitrary non-negative value.
In this paper, we will restrict ourselves to the charged case, 
i.e., $\mu_0 > 0$.
  
As we claimed in \cite{Kobayashi:2004ay}, 
the parametrization 
by $(c_1, c_2, r_0 )$ is not suitable to take the BPS limit, 
and so we defined a set 
of parameters $(c_1, \mu_0, \nu)$ which was defined by 
%--------------------------------------%
\begin{equation}
 r_0^{7-p} \equiv \frac{\nu \mu_0}{2k},\qquad 
 c_2^2-1 \equiv \frac{1}{\nu^2}.
 \label{new-para}
\end{equation} 
%--------------------------------------% 
The domain of $\nu$ becomes 
\begin{eqnarray}
\begin{cases}
0 \le \nu < \infty &\mbox{for the branch}\ I_+ , \\ 
-\infty < \nu \le 0 &\mbox{for the branch}\ I_-, 
\end{cases}
\end{eqnarray} 
since we choose $c_1 \ge 0, \ k \ge 0$ and $\mu_0 > 0$. 
Using these parameters, the RR-charge, the ADM mass 
and the dilaton charge are rewritten as
\begin{align}
 |Q| &= N_p \mu_0, \\
 M &= \left(\sqrt{1+{\nu^2}}
                  +\frac{3-p}{4}\frac{c_1 \nu}{k} \right) N_p \mu_0, \\
 D &= \left(\frac{3-p}{4}\sqrt{1+{\nu^2}}
          -\frac{(p+1)(7-p)}{4}\frac{c_1 \nu}{k} \right) N_p \mu_0
      = \frac{3-p}{4}M - \frac{c_1 \nu}{k}N_p \mu_0. 
\label{new-QM}
\end{align}

Then, it is easy to show that the three-parameter solution
(\ref{sln-1})--(\ref{sln-4})
reduces to the BPS black $p$-brane solution 
(\ref{extreme-1})--(\ref{extreme-3})
in the limit $\nu \rightarrow 0$ 
for arbitrary values of $c_1$ and $\mu_0$. 
Since $|Q|$ is unchanged in this limit, we can set  
$\mu_0$ to the same value as the black $p$-brane% 
\footnote{
For the neutral case, i.e., $|Q|=0$, another parametrization 
is suitable.
}.
We emphasize here that since there are three parameters, 
the BPS limit consists of two extremal limits, 
i.e.  
$M=|Q|$ and $D=\frac{3-p}{4}M$. 
The former is the ordinary BPS condition, which 
guarantees the preservation of the spacetime supersymmetry. 
Although the latter says that the dilaton charge is proportional 
to the ADM mass, its physical meaning is obscure at this stage. 
Nevertheless, if we take the BPS limit as a starting point, 
it is quite natural to characterize the solution by two 
non-extremality parameters, that represent the deviation from 
the BPS black $p$-brane solution.

From this observation, we introduce following quantities $m$ and $d$ 
defined by
\begin{align}
\label{def-m}
m &= \sqrt{1+\nu^2}, \\
\label{def-d}
d &= \frac{c_1 \nu}{k},
\end{align}
which are related to the quantities above as
\begin{align}
|Q| &= N_p \mu_0, \\
 M &= \left(m+\frac{3-p}{4}d \right) N_p \mu_0, \\
 D &= \left( \frac{3-p}{4}m -\frac{(p+1)(7-p)}{16}d \right)N_p \mu_0.
\end{align}
The region of these quantities are determined as 
\begin{eqnarray}
\begin{cases}
m \ge 1,\ d \ge 0 &\mbox{for the branch} \ I_+, \\ 
m \ge 1,\ d \le 0 &\mbox{for the branch} \ I_-. 
\end{cases}
\end{eqnarray} 
The BPS solution lies in $m=1$ and $d=0$ in the overlap of $I_+$ and $I_-$.
Note that the ADM mass and the dilaton charge are not conserved 
Noether charges and are frame-dependent quantities. 
Although their definition is ambiguous in some sense,  
they can characterize the deviation from the Minkowski space. 
On the other hand, our quantities $m$ and $d$ represent the deviation from 
the BPS black $p$-brane solution.
As opposed to the ADM mass $M$, $m$ is always greater than $1$. 
Hence, $m$ can be considered as a non-extremality parameter 
in the ordinary sense, while $d$ 
denotes the difference between the ADM mass $M$ and the dilaton 
charge $D$. 
Note that $d$ is related to the parameter $c_1$ directly. 
As we have shown in \cite{Kobayashi:2004ay}, 
the three-parameter solution with $c_1 =0$ denotes the 
spacetime which is produced 
from the D$p$\={D}$p$-brane system with a constant tachyon VEV.
Since a main purpose of this paper 
is to clarify the stringy meaning of $c_1$, 
our task is to find sources in the string theory which produces 
$d \neq 0$ structure.

At the end of this section, we give the long distance behavior 
of the three-parameter solution (\ref{sln-1})--(\ref{sln-4}), 
which is given by the leading term of $1/r$ expansion.
In terms of $(\mu_0, m, d)$ it is given by
\begin{align}
\label{asymp-new-1}
 e^{2A(r)} &= 1-\frac{7-p}{8}\left(m+\frac{3-p}{4}d\right)\frac{\mu_0}{r^{7-p}}
         +{\cal O}\left(1/r^{2(7-p)}\right),  \\
\label{asymp-new-2}
 e^{2B(r)} &= 1
  +\frac{p+1}{8}\left(m+\frac{3-p}{4}d\right)\frac{\mu_0}{r^{7-p}}
  + {\cal O}\left(1/r^{2(7-p)}\right),  \\
\label{asymp-new-3} 
\phi(r) &= \left( \frac{3-p}{4}m - \frac{(p+1)(7-p)}{16}d \right)
           \frac{\mu_0}{r^{7-p}}
  + {\cal O}\left(1/r^{2(7-p)}\right),  \\
 e^{\Lambda(r)} &= \eta \frac{\mu_0}{r^{7-p}}
  + {\cal O}\left(1/r^{2(7-p)}\right). 
\label{asymp-new-4}
\end{align}
We will compare them with the massless emission from the source 
in the string theory in the next section.

%%%%%%%%%%%%%%%%%%%%%%%%%%%%%%%%%
\section{Excitations on the D$p$\={D}$p$-brane System}
\label{sec:string-side}

In this section, we investigate stringy counterparts of the three-parameter 
solution. 
We argue that the geometry expressed by the three-parameter solution is
made by D-brane systems which are in general sources of closed strings in the
bulk. 
We stress that we only consider static sources 
and do not take into account interactions of open strings on the D-brane. 
Therefore they can be consistent sources of supergravity fields, 
even if they are unstable system. 
As a first step, we deform the boundary state for BPS D$p$-branes 
appropriately, 
and compare the long distance behavior of the solution with the massless 
emissions from them.
And then, we give some examples.

\subsection{Deformations of Boundary States}
\label{sec:ABC}

To obtain the three-parameter solution, we have required that the
solution has the symmetry 
$ISO(1,p)\times SO(9-p)$ and it carries a RR $(p+1)$-form charge 
as explained in the previous section. 
Therefore 
the source of closed strings, which produces the three-parameter solution,
should respect the following restrictions at least in the low energy
regime; 
\begin{align}
\label{ansatz}
1)\quad  & \text{has the symmetry $ISO(1,p)\times SO(9-p)$}, \nn  
2)\quad  & \text{carries only the RR $(p+1)$-form charge,} \\
3)\quad  & \text{has the $\delta$-function distribution in the transverse space.} 
\nonumber 
\end{align}
Note that the third condition can be slightly relaxed (see below). 

Needless to say, the system of coincident BPS D$p$-branes is such an example. 
$N$ BPS D$p$-branes are expressed by the boundary state, 
\begin{equation}
\label{GSO projected}
\ket{Dp}= {\cal P}_{\rm GSO}\Bigl(
  N\ket{Bp}_{\rm NS}
 +N\ket{Bp}_{\rm RR}\Bigr), 
\end{equation}
with%
\footnote{
In this expression, we have omitted the label ``$+$'' of the 
spin structure. 
} 
\begin{equation}
\ket{Bp}_{\rm NS(RR)}
= \frac{T_p}{2} \exp\left[
  {-\sum_{n=1}^{\infty} 
  \frac{1}{n} \alpha_{-n}^M S_{MN} \talpha_{-n}^N} 
  +i\sum_{r>0}^{\infty} b_{-r}^M S_{MN} \tb_{-r}^N
 \right] \ket{p_\mu=0,x^i=0}_{\rm NS(RR)},   
\label{BPS boundary state}
\end{equation}
where $T_p$ is the tension for a single D$p$-brane, 
$\mu=0,1,\cdots,p$ are the directions longitudinal to the worldvolume, 
$i=p+1,\cdots,9$ are the directions transverse to the D$p$-branes, 
and 
$\alpha_{-n}^M$ and $b_{-r}^M$ ($\talpha_{-n}^M$ and $\tb_{-r}^M$) are the 
creation operators of the modes of the left-moving (right-moving)
worldsheet bosons and fermions, respectively.
$S_{MN} = {\rm diag} \left(\eta_{\mu\nu}, -\delta_{ij} \right)$ gives 
Neumann (Dirichlet) boundary conditions on the string worldsheet 
in $\mu (i)$ directions, respectively.
Here the number of the D$p$-branes $N$ is the only microscopic parameter.
In this case, it is shown that 
massless emission from this boundary state agrees with 
the long distance behavior of the BPS black $p$-brane \cite{DiVecchia:1997pr}
under the identification,  
\begin{eqnarray}
\mu_0
= \frac{2\kappa N T_p}{(7-p) \omega_{8-p}}, 
\label{RR-relation}
\end{eqnarray}
which relates the macroscopic parameter $\mu_0$ 
with the microscopic parameter $N$. 
Of course, the system with $\bar{N}$ $\bar{\rm D}p$-brane also satisfies
the same ansatz, which is simply given by replacing $N$ to $-\bar{N}$ in the 
RR-sector of (\ref{GSO projected})%
\footnote{
In this case, (\ref{RR-relation}) is same but $\eta$ replaced by $-1$. 
}.
We can also consider the system with $N$ D$p$-branes and $\bN$ \=D$p$-branes
without tachyon field which is given by the linear combination of 
the both above. 

We now take the BPS boundary state (\ref{GSO projected}) as the starting point 
as mentioned in Sec.1 and  
consider possible deformations of it 
with keeping the ansatzes (\ref{ansatz}), 
which should be the counterpart of the discussion 
in Sec.2, that is, on the supergravity side. 
One might try to deform it with massless open string
excitations on the brane. 
However, the symmetry $ISO(1,p)\times SO(9-p)$ prevents
the fluctuation of massless scalar fields, 
and there should be no gauge flux since otherwise 
it produces other NSNS/RR-charges. 
Therefore, the excitations 
on the branes should be at least tachyonic and/or massive%
\footnote{
Strictly speaking, the linear combination 
of the boundary state and other closed string states is also
possible, but we do not consider this option.
}.
%We will not try to find the most general form of the deformation but study
%some minimal choices, whose effects give non-trivial contributions to 
%$c_1$. 

In order to make the discussion transparent, 
we formally deform the boundary state as follows. 
In the NSNS-sector, we set
%-------------------------------------------------------%
\begin{equation}
\ket{B_p'}_{NS} = C N \frac{T_p}{2}  
            \exp \left[
                 -\sum_{n=1}^{\infty} 
                  \frac{1}{n}\alpha_{-n}^{M} S_{MN}^{(n)} \talpha_{-n}^N
                  +i\sum_{r>0} b_{-r}^{M} S_{MN}^{(r)} b_{-r}^N 
                 \right]  
            \ket{p_{\mu}=0, x^i=0}_{NS},
\label{general boundary state}
\end{equation}
%--------------------------------------------%
where
%------------------------------------------%
\begin{equation}
S_{MN}^{(n,r)} = {\rm diag} 
\left(A^{(n,r)}\eta_{\mu\nu}, -B^{(n,r)}\delta_{ij} \right).  
\end{equation}
%------------------------------------------%
Here $A^{(n,r)}$, $B^{(n,r)}$ and $C$ are deformation parameters which deform the boundary 
state without changing its symmetry. 
When we take $A^{(n,r)}=B^{(n,r)}=C=1$, 
the boundary state (\ref{BPS boundary state}) for the $N$ BPS D$p$-branes
is reproduced. 
We also deform the RR-sector of the boundary state (\ref{BPS boundary state})
as the same manner but $C=1$ in this case
since we fix the RR-charge to be the same as the $N$ D$p$-branes. 
We give some comments in order. 
First, one might assume the source object is 
a $\delta$-function type located at the origin in the transverse space.
But it will turn out to be too restrictive and can be relaxed slightly: 
the ``width'' of the source allowed to be less than the string
length. 
We will come back to this point in the next subsection.  
Second, the effect of $(A^{(n,r)}, B^{(n,r)})$ in $S_{MN}^{(n,r)}$ 
is the deformation of the boundary condition. 
Here we note that only the $r=1/2$ mode contributes the massless mode of 
closed strings emitted from the boundary state. 
In other words, the deformations for other modes, 
$S_{MN}^{(n)}$ and $S_{MN}^{(r\ne 1/2)}$, 
are arbitrary in our analysis below. 
However, they should be related among them by other consistency, 
such as the modular invariance 
although we do not discuss them at this stage. 
Third, $C$ denotes the difference of the overall normalization in the 
NSNS-sector from that in the RR-sector.
For example, in the system of $(N+M)$ D$p$-branes and $M$ $\bar{\rm D}p$-branes, 
we take $CN=N+2M$, the total number of branes, 
while the RR-sector has a fixed charge proportional to $N$.

Then, we can calculate the massless emissions with momentum $p_M$ 
from this deformed state as we have done in \cite{Kobayashi:2004ay}. 
In the NSNS-sector 
there are the graviton $h_{MN}$ and the dilaton $\phi$, that are
\begin{eqnarray}
f(p)&=&\bra{f(p)}\Delta\ket{D^{\prime}p}
 =-CN\frac{T_p}{2} \frac{V_{p+1}}{p_i^2} \epsilon_{MN}^f S^{(1/2)MN},
\end{eqnarray}
where $\ket{D^{\prime}p}$ denotes the GSO projected deformed state, 
$\Delta$ is the closed string propagator and $\epsilon_{MN}^f$ is 
the appropriate polarization tensor for $f=h_{MN},\phi$ 
\cite{Kobayashi:2004ay}.
There is a similar expression for the RR $(p+1)$-form $e^{\Lambda}$. 
In this case, since we set $C=1$ for the RR-sector and since 
only the zero-mode part contributes to the massless fields, 
we obtain same result as the BPS D$p$-brane.
After the Fourier transformation to the position space, we obtain  
%--------------------------------------------%
\begin{align}
\label{graviton-new}
h_{MN}(r) 
 &= \left(-\frac{7-p}{8}\eta_{\mu\nu}, \frac{p+1}{8}\delta{ij} \right)
    \frac{C(A+B)}{2}\frac{2\kappa N T_p}{(7-p)\omega_{8-p}}\frac{1}{r^{7-p}}, \\
\label{dilaton-new}
\phi(r)
 &= \frac{C}{8}\left[-(p+1)A+(7-p)B \right]
 \frac{2\kappa N T_p}{(7-p)\omega_{8-p}}\frac{1}{r^{7-p}}, \\
\label{ramond-new}
e^{\Lambda}(r)
 &= \pm \frac{2\kappa N T_p}{(7-p)\omega_{8-p}}\frac{1}{r^{7-p}},
\end{align}
where we write $A=A^{(1/2)}$ and $B=B^{(1/2)}$.
%--------------------------------------------%
Comparing these quantities (\ref{graviton-new})-(\ref{ramond-new}) 
with the long distance behaviors
(\ref{asymp-new-1})-(\ref{asymp-new-4}), we obtain the relation 
between the macroscopic quantities and the microscopic ones as 
\begin{align}
& \mu_0 = \frac{2\kappa N T_p}{(7-p) \omega_{8-p}}, \\
\label{m-ABC}
& m = \frac{C}{8}\left[(p+1)A+(7-p)B\right], \\
\label{d-ABC}
& d = \frac{C}{2}(A-B).
\end{align}
We see from the eqs. (\ref{m-ABC}) and (\ref{d-ABC}) that 
the two non-extremality parameters $m$ and $d$ in the three-parameter solution
correspond to the deformation parameter $(A,B,C)$ 
on the string theory side.
The first non-extremality $m\ne 1$ can be realized even if $A=B=1$, 
by taking $C\ne 1$. 
It means that the breaking of the BPS condition can occur 
within the ordinary boundary condition.
However, the second non-extremality $d \neq 0$ requires 
necessarily $A \neq B$. 
Therefore, in order to have the three-parameter solution with nonzero $c_1$, 
it is necessary to deform the boundary condition 
from the ordinary one of the BPS D$p$-branes so as to satisfy $A \neq B$. 
In the next subsection, 
we give examples of such stringy objects.

\subsection{Examples}

In this subsection, we give three typical examples realizing 
the deformed states in the string theory 
which we referred in the previous subsection by considering 
various types of the tachyon condensation. 
The first example is the case considered in  
\cite{Kobayashi:2004ay}. 
The other two examples are obtained 
from the tachyon condensation from higher/lower dimensional 
unstable D-brane systems.
In general, deformations are interpreted as 
tachyonic/massive excitations on the D$p$\={D}$p$ system.

\subsubsection{Case 1: D$p$\={D}$p$ system with a tachyon VEV}
 
The first example is the D$p$\={D}$p$ system with a constant tachyon profile.
To be precise we consider 
the $(N+M)$ D$p$-branes and $M$ $\bar{\rm D}p$-branes.
The gauge symmetry of the  worldvolume theory is $U(N+M)\times U(M)$
and there is a complex tachyon field $T(x)$ in the 
bi-fundamental representation of the gauge group. 
Here we consider the case where the $M$ D$p$\=D$p$-pairs vanish
and $N$ D$p$-branes remain. 
Namely, we decompose the $M \times (N+M)$ matrix by $M \times N$ and 
$M \times M$ components and set the tachyon profile as
\begin{eqnarray}
T(x)=\left(
\begin{array}{c}
0\\
\hline
t
\end{array}
\right),
\end{eqnarray}
where $t$ is a constant $M \times M$ matrix.
Note that other open string excitations, such as 
gauge fields, massless scalar fields and a non-constant tachyon,   
break the symmetry $ISO(1,p)\times SO(9-p)$.

It is known that this system can be expressed by the 
off-shell boundary state 
on which the boundary interaction for the tachyon field 
is turned \cite{Takayanagi:2000rz,Asakawa:2002ui}. 
For our case, the NSNS-sector of the boundary state 
is then given by 
%----------------------------------------------%
 \begin{align}
\ket{B_p';t}_{\rm NS} 
&= 
\left[
N + 2{\rm Tr}_{M}\, e^{-|t|^2}
\right] \frac{T_p}{2}\,
e^{-\sum_{n=1}^{\infty} \frac{1}{n}
   \alpha_{-n}^{M}S_{MN}\talpha_{-n}^{N}
   +i\sum_{r>0} b_{-r}^{M} S_{MN}
   \tb_{-r}^{N}}
   \ket{p_\mu =0, x^i =0}_{\rm NS},
\label{boundary state}
\end{align}
%----------------------------------------------%
where $S_{MN} = (\eta_{\mu\nu}, -\delta_{ij})$ and 
the trace is taken over $M \times M$ matrices. 
Note that it is the same as (\ref{BPS boundary state}) except for the 
overall factor, $\left[N + 2{\rm Tr}_{M}\, e^{-|t|^2}\right]$,  
which simply comes from the tachyon potential $V(T)\sim {\rm Tr}e^{-|T|^2}$. 
It reduces to the $N+2M$ in the limit $|t|\rightarrow 0$, corresponding to the 
$(N+M)$ D$p$-branes and $M$ $\bar{\rm D}p$-branes, 
while reduces to $N$ in the limit $|t|\rightarrow \infty$,
corresponding to the $N$ D$p$-branes. 

{}From this state, we can read off $(A, B, C)$ by comparing 
(\ref{boundary state}) with (\ref{general boundary state}) as 
\begin{equation}
A=B=1,\qquad \ C = 1 +\frac{2}{N}\ {\rm Tr}_M \ e^{-|t|^2}. 
\end{equation}
Therefore, macroscopic quantities $m$ and $d$ are given by
\begin{equation}
m=1 +\frac{2}{N}\ {\rm Tr}_M \ e^{-|t|^2},\qquad \ d=0.
\end{equation}
We see that the constant tachyon $t$ only contributes to the ADM mass.
Therefore, this system corresponds to at most two-parameter subset of the full 
three-parameter solution as we mentioned in \cite{Kobayashi:2004ay}, 
but it is an trivial example 
for our present purpose.

\subsubsection{Case 2: Unstable D9-branes to BPS D$p$-branes}

Next example is the tachyon condensation 
from unstable D$9$-branes to the BPS D$p$-branes. 
For simplicity, we consider Type IIB superstring theory, 
that is, the tachyon condensation from pairs of D9-branes and
\=D9-branes to BPS D$p$-branes. 
For Type IIA superstring theory, we start from non-BPS D9-branes  
and the discussion is completely parallel. 
Since a single BPS D$p$-brane is obtained from a pair of 
D(p+2)-brane and \=D(p+2)-brane, 
it is clear that it is obtained from $2^{\frac{7-p}{2}}$ 
pairs of D9-brane and \={D}9-brane via the tachyon condensation.  
Then let us consider $N \times 2^{\frac{7-p}{2}}$ pairs of D9\={D}9-branes 
and the tachyon profile, 
\begin{equation}
T(x)=u \gamma_i x^i \otimes 1_N,
\label{D9 tachyon}
\end{equation}
where $i=p+1,\cdots,9$ are the transverse directions 
to the resulting D$p$-brane,
$\gamma_i$ are the $SO(9-p)$ gamma matrices 
and $u$ is a real parameter of dimension of mass. 
It is known that this system reduces to $N$ BPS D$p$-branes 
in the limit $u\rightarrow \infty$%
\footnote{For details we refer the literatures, which discussed  
 in the context of the worldvolume field theory \cite{Witten:1998cd},  
of the boundary string field theory 
\cite{Takayanagi:2000rz}\cite{Kraus:2000nj}
and of the boundary state \cite{Asakawa:2002ui}.
}.
If we replace $T$ to $-T$, the final state becomes \={D}$p$-branes.

Here the effect of the tachyon profile (\ref{D9 tachyon}) 
with $u\ne 0$ is twofold:
First, it breaks the original global symmetry $ISO(1,9)$ down to 
$ISO(1,p)\times SO(9-p)$. 
Next, it produces the correct 
RR-charge of $N$ D$p$-branes which is independent of the value of $u$. 
Therefore, this system satisfies the first two items of the ansatzes
(\ref{ansatz}) for an arbitrary value of $u\ne 0$. 
However, accurately speaking, it does not satisfy the third item. 
In fact, 
the intermediate state ($0<u< \infty$) is neither D$9$-branes nor D$p$-branes
but its energy density in the transverse direction 
has the Gaussian-shaped distribution with a width $\sim 1/u$. 
(This is easily seen by noting the tachyon potential has the form 
$V(T)\sim e^{-u^2 x^ix_i}$.) 
Therefore, if we allow the width of the source less than the string
length, $u$ should be sufficiently larger than $1/\sqrt{\alpha'}$. 
We will evaluate it more precisely below. 

This system is again described by the off-shell boundary 
state of the D9\={D}9 system with the tachyon  profile (\ref{D9 tachyon})
turned on \cite{Asakawa:2002ui}. 
It is given by in the NSNS-sector,  
\begin{equation}
\ket{Bp';u}_{\rm NS}
=\frac{T_9}{2} \int [d\bX^{M}] 
{\Tr} \widehat{\rm P} e^{\int d\hsigma \bM(\hsigma)}
\ket{\bX^{M}}_{\rm NS}, 
\end{equation}
with 
\begin{equation}
\bM=\left(
\begin{array}{cc}
0 & T(\bX) \\ 
T^\dagger(\bX) & 0
\end{array}
\right), 
\label{original}
\end{equation}
where $T_9$ is the tension of a single D$9$-brane, 
$T(x)$ is given by (\ref{D9 tachyon}) and the trace is taken over 
the Chan-Paton space of size $2^{\frac{9-p}{2}}N$. 
It represents the Neumann boundary state 
with arbitrary numbers of tachyon vertex operators attached. 
By evaluating the path integral and rewriting it with oscillators, 
we obtain (see Appendix for more details)
\begin{align}
\ket{Bp';u}_{\rm NS}
=&N\frac{T_p}{2} \,\ e^{-\frac{1}{4u^2} \int \bP_i D \bP_i}
\ket{\bP_\mu =0,\bX^i =0}_{\rm NS}
\label{transverse massive}\\
=&N\frac{T_p}{2}
F(y)^{9-p} 
e^{
   -\sum_{n=1}^{\infty} \frac{1}{n}
   \alpha_{-n}^{M}S_{MN}^{(n)}\talpha_{-n}^{N}
   +i\sum_{r>0} b_{-r}^{M}S_{MN}^{(r)}
   \tb_{-r}^{N}} 
   \int dp_i e^{-\frac{p_i^2}{8\pi u^2}}
   \ket{p_\mu =0, p_i}_{\rm NS}, 
\label{gaussian}
\end{align}
where $y \equiv 4\pi \alpha' u^2$ is a dimensionless parameter, 
and the function $F(y)$ is given by  
\begin{equation}
F(y) \equiv \frac{4^y y^{1/2}\Gamma^2(y)}{\sqrt{4\pi}\Gamma(2y)}, 
\end{equation}
and 
\begin{equation}
 S_{MN}^{(n)} = 
\left(\eta_{\mu\nu},-\frac{y-n}{y+n}\delta_{ij}\right), 
\qquad 
 S_{MN}^{(r)} = 
\left(\eta_{\mu\nu},-\frac{y-r}{y+r}\delta_{ij}\right).  
\end{equation} 
The corresponding state for the RR-sector is similarly obtained.

\begin{figure}[t]
  \begin{center}
    \includegraphics[scale=1]{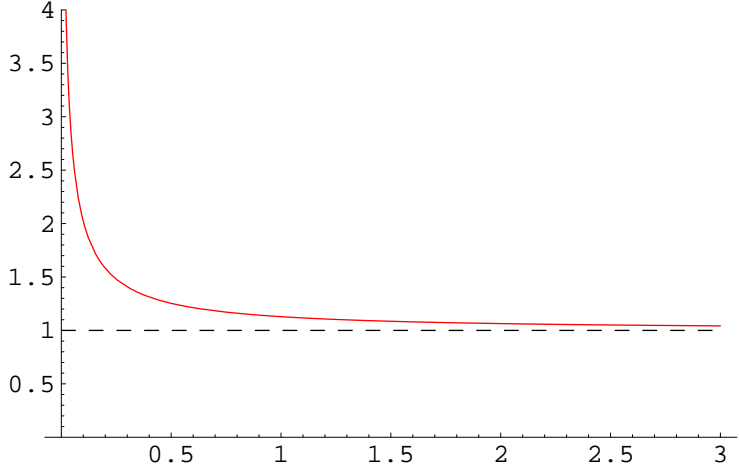}
  \end{center}
  \caption{Profile of $F(y)$. The behavior of this function is the same
 as that of $C$, that is, $C\to\infty$ for $y\to0$ and $C\to 1$ for
 $y\to\infty$.} 
  \label{fig:F.eps}
\end{figure}

It is instructive to see the behavior of this state under the 
change of the value $u$ $(0 < u < \infty)$. 
First of all, 
we can identify the state (\ref{gaussian}) 
with the deformed state (\ref{general boundary state}) 
(apart from the zero-mode part) as
\begin{equation}
A^{(n,r)}=1, \qquad  B^{(n)}=\frac{y-n}{y+n}, \qquad 
B^{(r)}=\frac{y-r}{y+r}, \qquad C=F(y)^{9-p}.
\end{equation}
$A^{(n,r)}$ is independent of $y$, which means that 
the tangential directions always satisfy the Neumann boundary condition.
On the other hand, $B^{(n,r)}$ is a monotonically increasing function of
$y$, starting from $-1$ (Neumann) at $y=0$ and goes to $1$ (Dirichlet)
 at $y\rightarrow\infty$.
It means that the transverse directions satisfy a mixed boundary
condition of Dirichlet and Neumann. 
Roughly speaking, the boundary condition is Dirichlet-like for
$B^{(n,r)}<0$ and Neumann-like for $B^{(n,r)}>0$. 
Probing only by the massless fields, 
this state has similar feature as the BPS D$p$-branes at least 
in the range sufficiently larger than $y=1/2$, 
although other higher modes can be Neumann-like in this region. 
$C$ behaves as a monotonically decreasing function of $y$ 
with $C \rightarrow \infty$ for $y\rightarrow 0$ 
and $C \rightarrow 1$ for $y\rightarrow \infty$. 
(We give the profile of the function $F(y)$ in Fig.\ref{fig:F.eps}.) 
Therefore the effective tension is always greater than that of the BPS
D$p$-brane. 
Although this factor itself is not well-behaved at $y\rightarrow 0$, 
combined with the contribution from the zero-mode, 
we recover the correct tension for the D$9$-brane (see Appendix).
The zero-mode part depends directly on $u$ and means the  
Gaussian-shaped distribution in the transverse with a $({\rm width})^2 = 1/{8\pi u^2}$. 
Strictly speaking, this is excluded 
from our assumption for the deformed state. 
However, since we only concern the massless emission, it is enough to 
demand the delta-function source at the low energy, that is, the value of $u$ 
such that $1/{8\pi u^2} \lesssim  \alpha' $ is in fact well-localized 
in the limit $\alpha' \rightarrow 0$. 
This condition is equivalent to $y \gtrsim 1/2$, 
which is the same region discussed previously.
When the width is the order of the string scale, then the supergravity 
approximation is no longer valid and the $\alpha'$-correction in the 
equation of motion and also the source should be taken into account.
In this case, the zero-mode factor in (\ref{gaussian}) is nothing but the 
$\alpha'$-corrected $\delta$-function source considered 
in \cite{Tseytlin:1995uq} in the context of the smearing of the singularity.

\begin{figure}[t]
  \begin{center}
    \includegraphics{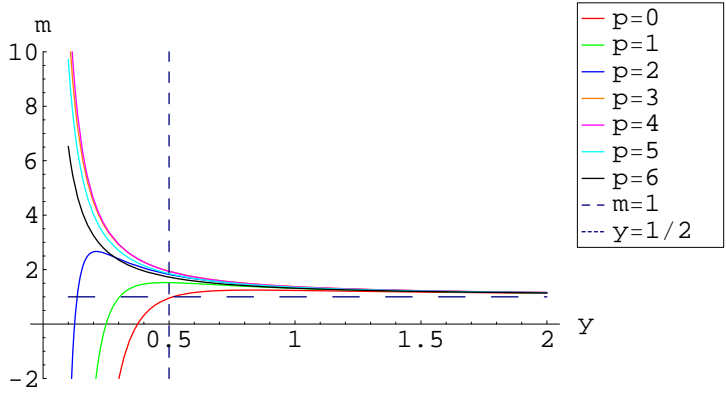}
  \end{center}
  \caption{
 Profiles of $m$'s of the unstable D9-branes for various $p$'s.
 For the region $y \gtrsim 1/2$, $m$ is very closed to 1, that is, 
 the solution is near extremal in the whole parameter space of the 
 three-parameter solution.
 }
  \label{fig:m_D9.eps}
\end{figure}

\begin{figure}[t]
  \begin{center}
    \includegraphics{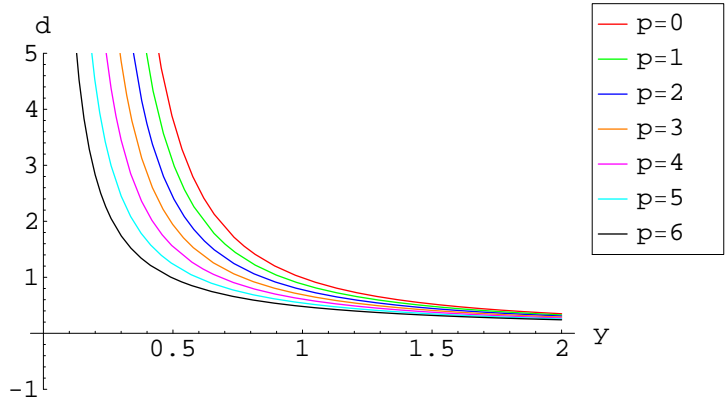}
  \end{center}
  \caption{
 Profiles of $d$'s of the unstable D9-branes for various $p$'s. 
}
  \label{fig:d_D9.eps}
\end{figure}

Thus, at least in the region $y \gtrsim 1/2$, 
we can identify this microscopic state with the macroscopic quantities as
\begin{equation}
m=F(y)^{9-p}\frac{8y+p-3}{4(2y+1)}, \qquad  d=F(y)^{9-p}\frac{1}{2y+1}.
\end{equation}
The behavior of $m$ and that of $d$ as function of $y$ are shown 
in Fig.\ref{fig:m_D9.eps} 
and Fig.\ref{fig:d_D9.eps}, respectively.
Then, we see that they represent a trajectory in the $(m,d)$ space 
and belong to the branch $I_{+}$: $m \ge 1$, $d\ge 0$.
This gives two-parameter subset of the three-parameter solution, 
which possesses an non-extremal dilaton charge.

\begin{figure}[t]
\begin{center}
\includegraphics[scale=0.6]{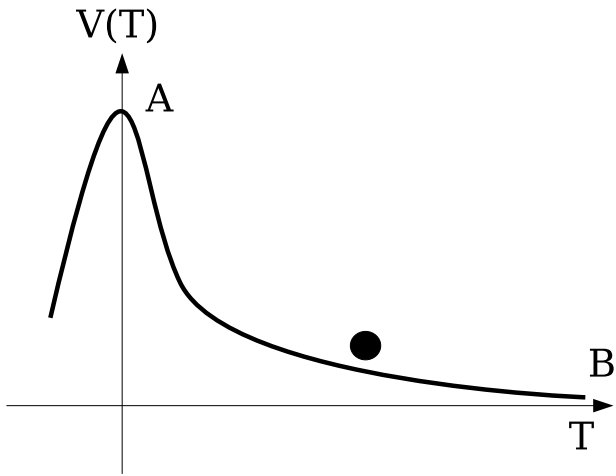}
\end{center}
\caption{
The black circle corresponds to the state in the way of the tachyon
 condensation from a system of unstable D9-branes to that of BPS D-branes. 
{}From the viewpoint of the (false) vacuum (A), it can be regarded as a
 tachyon excited state. 
On the other hand, if we look it at from the (true) vacuum (B), 
it can be regarded as massive excited state. 
}
\label{tachyon-massive}
\end{figure}

As a final remark, we note that the deformation here is 
also regarded as a massive excitation of open strings 
on the BPS D$p$-branes, 
since a tachyonic excitation on the top of the potential 
is equivalent to the massive excitation from the viewpoint 
of the bottom (See Fig.\ref{tachyon-massive}.).   
In fact, the original expression (\ref{original}) 
represents the deformation of the D9\={D}9 system 
by a tachyonic excitation,  
while in (\ref{transverse massive}) the same system is 
described by the D$p$-branes with massive vertex operators 
$\int d\widehat \sigma \bP_i D \bP_i$ turned on.

\subsubsection{Case 3: Unstable D-instantons to BPS D$p$-branes}

The third example is the tachyon condensation from 
a system of unstable D-instantons to the BPS D$p$-branes. 
In this case, the BPS D$p$-branes can be constructed as bound states of
infinitely many unstable D-instantons%   
\footnote{For details we refer the literatures discussed that issue 
 from the viewpoint of the worldvolume theory 
\cite{Terashima:2001jc}\cite{Kluson:2000gr}\cite{Asakawa:2001vm}
and from the viewpoint of the boundary state \cite{Asakawa:2002ui}.
}.
For the concreteness, we consider a system of non-BPS D-instantons  
in Type IIA string theory.
There are ten scalar fields $\Phi^M$ and a tachyon field $T$ on them.
They are regarded as self-adjoint operators acting on the infinite dimensional 
Hilbert space, since the worldvolume is zero-dimensional and the matrix-size 
is infinite.
Then the configuration representing $N$ BPS D$p$-branes is given by 
\begin{gather}
 T = v \hp_\mu \otimes \gamma^\mu \otimes 1_N , 
\label{Dp solution}\\
 \Phi^\mu = \hx^\mu \otimes 1 \otimes 1_N , \qquad 
 \Phi^i =0, 
\end{gather}
where $\hx^\mu$ and $\hp_\mu$ are operators on a Hilbert space 
satisfying the canonical commutation relation, 
\begin{equation}
 \left[\hx^\mu,\hp_\nu\right]
 = i\delta^\mu_{\ \nu},
\end{equation}
$\mu=0,1,\cdots,p$ are the worldvolume direction and 
$i=p+1,\cdots,9$ are the transverse direction and  
$\gamma^\mu$ are Hermitian gamma matrices. 
$v$ is a real parameter with dimension of length and 
this configuration becomes an exact solution in the limit 
$v\rightarrow \infty$.

The intermediate state $(0<v<\infty)$ can be interpreted as a 
$(p+1)$-dimensional object, which has a fuzzy worldvolume in some sense. 
This is most easily understood as follows.
Recall that the each set of eigenvalues 
of the scalar fields $\{\Phi^M\}$ 
represents the position of each individual D-instanton. 
In the absence of the tachyon profile $(v=0)$, 
they are distributed uniformly on the $(p+1)$-dimensional plane,
since the spectrum of $\hx^\mu$ spans the real axis. 
Note that they are still the collection of the D-instantons.
However, if we turn on the tachyon profile $(v>0)$, 
D-instantons become correlated among them.
As seen from the tachyon potential 
$V(T)\sim {\rm Tr} e^{-v^2 \hp_\alpha \hp^\alpha}$,
the momentum distribution is localized around the origin 
of the momentum space with a width $\Delta p\sim 1/v$. 
This means (in an appropriate way) 
that the position of each D-instanton becomes uncertain 
with an amount of $\Delta x \sim v$.
Then, in the limit $v\rightarrow \infty$, 
it becomes $\Delta x =\infty$ and 
we cannot observe the individual D-instantons 
and the worldvolume of D$p$-brane appears.

The off-shell boundary state describing this system is given by 
the Dirichlet boundary state with the scalar and tachyon fields 
turned on as the boundary interaction. 
In the NSNS-sector, after calculating the trace over the Chan-Paton
Hilbert space, we have 
\begin{align}
\ket{Bp';v}_{\rm NS} = &
N\frac{T_p}{2} \int \left[d\bX^\mu \right] 
 e^{-\frac{1}{4v^2}\int D\bX^\mu D^2\bX^\mu}
 \ket{\bX^\mu,\bX^i=0}_{\rm NS},
\label{longitudinal massive}\\
=& N\frac{T_p}{2} F(y)^{p+1} 
e^{
   -\sum_{n=1}^{\infty} \frac{1}{n}
   \alpha_{-n}^{M}S_{MN}^{(n)}\talpha_{-n}^{N}
   +i\sum_{r>0} b_{-r}^{M}S_{MN}^{(r)}
   \tb_{-r}^{N}} 
   \ket{p_\mu =0, x^i=0}_{\rm NS},
\label{haussian}
\end{align}
where $y$ is a dimensionless parameter 
$y \equiv {v^2}/{\pi \alpha'}$, 
$F(y)$ is a function, 
\begin{equation}
F(y) \equiv \frac{4^y y^{1/2}\Gamma^2(y)}{\sqrt{4\pi}\Gamma(2y)},
\end{equation}
and 
\begin{equation}
 S_{MN}^{(m)} = 
\left(\frac{y-m}{y+m}\eta_{\mu\nu},-\delta_{ij}\right), 
\qquad 
 S_{MN}^{(r)} = 
\left(\frac{y-r}{y+r}\eta_{\mu\nu},-\delta_{ij}\right), 
\end{equation} 
which is very similar to (\ref{gaussian}). 
Then we can identify this state with the deformed state 
(\ref{general boundary state}) with setting the parameters as  
\begin{equation}
A^{(n)}=\frac{y-n}{y+n}, \qquad 
A^{(r)}=\frac{y-r}{y+r}, \qquad B^{(n,r)}=1, \qquad  C=F(y)^{p+1}. 
\end{equation}
Note that the zero-mode part is exactly the delta-function in this case. 
The boundary condition in the worldvolume direction is now 
deformed to $A^{(n,r)}$, which connects  
$A^{(n,r)}=-1$ (Dirichlet) for $y\rightarrow 0$ and 
$A^{(n,r)}=1$ (Neumann) for $y\rightarrow \infty$. 
This is understood more easily from the viewpoint of 
the deformation from the $N$ BPS D$p$-branes, 
that is, the expression 
(\ref{longitudinal massive}), on which a massive vertex operator
$\frac{1}{4v^2}\int D\bX^\mu D^2\bX_\mu$ turned 
\footnote{
See also recent discussions on the relevance of vertex operator here and 
fuzziness of the worldvolume \cite{Ellwood:2005yz}\cite{Terashima:2005ic}.
}. 
Since such a vertex operator makes the end point of the open string massive, 
the freely moving endpoint at $y\rightarrow \infty$ becomes 
heavier and heavier as $y$ decreases, 
and finally it is completely frozen in the limit of $y\to 0$. 
The behavior of $C$ as a function of $y$ is the same as the previous 
example. Here the divergence at $y=0$ is reflected simply by the original 
system has infinite number of non-BPS D-instantons.

\begin{figure}[t]
  \begin{center}
   \scalebox{.90}{\includegraphics{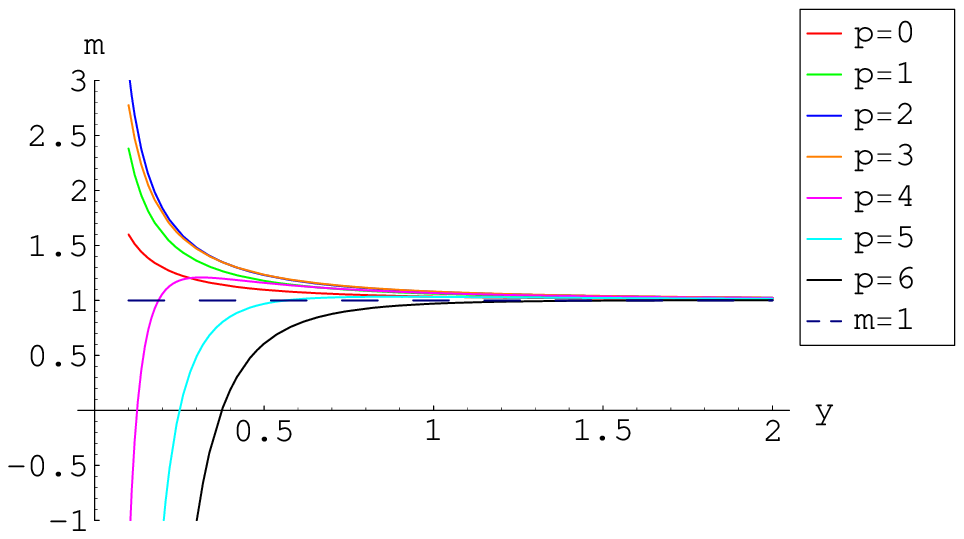}}
    \scalebox{.65}{\includegraphics{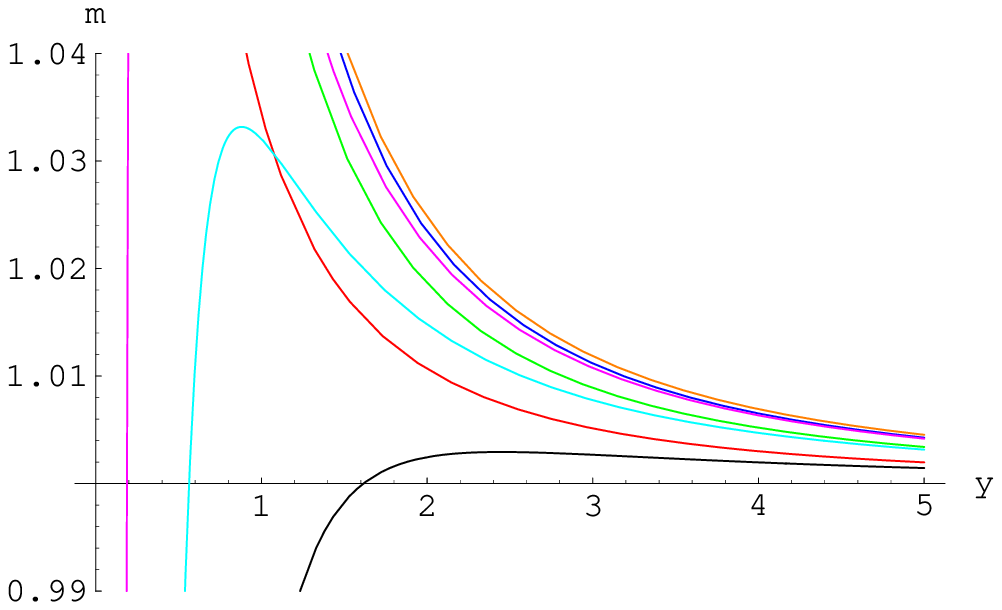}}
  \end{center}
  \caption{
 The left figure is the profiles of $m$'s of the system of unstable
 D-instantons for various $p$'s. 
  The right one is a close-up of the left one around $m= 1$. 
 From these figures, it is clear that all $m$'s are always larger than
 $1$ for sufficiently large $y$ 
 and asymptotically approach $1$ as $y$ increases.
 }
  \label{fig:m_Dinstanton.eps}
\end{figure}

\begin{figure}[t]
  \begin{center}
    \includegraphics{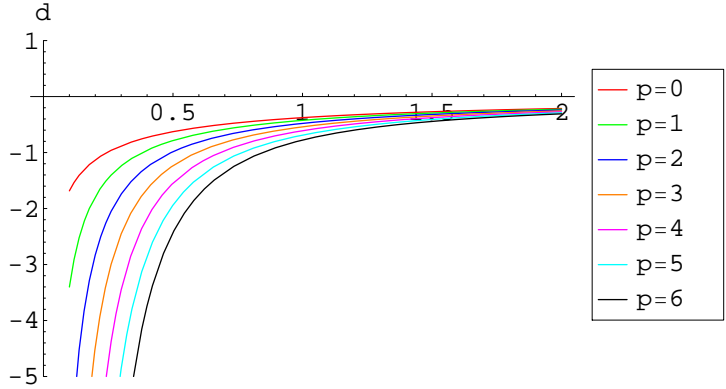}
  \end{center}
  \caption{
            Profiles of $d$'s of the system of unstable D-instantons for various $p$'s.
            All $d$'s takes arbitrary negative value. 
            }
  \label{fig:d_Dinstanton.eps}
\end{figure}

Comparing the massless emission from this state and the long range behavior 
of the three-parameter solution, we can relate the macroscopic to  
the microscopic quantities as
\begin{equation}
m=F(y)^{p+1}\frac{8y-p+3}{4(2y+1)}, \qquad 
d=-F(y)^{p+1}\frac{1}{2y+1}, 
\end{equation}
and the profiles of these parameters are depicted in 
Fig.\ref{fig:m_Dinstanton.eps} and Fig.\ref{fig:d_Dinstanton.eps}. 
In this case, one might conclude that the description of the
supergravity is valid for all region of $y$ since the configuration of
D-instantons does not seem to break the ansatzes (\ref{ansatz}). 
However, as shown in Fig.~\ref{fig:m_Dinstanton.eps}, 
$m$ is smaller than $1$ if $y$ is too small,  
which is out of the parameter region of the three-parameter solution. 
This means that this system is consistent 
with the three-parameter solution only when $y$ is sufficiently large. 
This would be understood as follows. 
As mentioned above, when $v$ $(y)$ is small, the boundary of open
strings on the ``D-brane'' is localized in the tangential directions. 
From the viewpoint of closed strings, this means that the size of loops
that construct the (deformed) boundary state is localized when $y$ is
small, that is, the translational invariance is spontaneously broken. 
This is also related to the uncertainty of the position of D-instantons 
in the tangential directions. 
In order to recover the translational invariance of closed string on the brane, 
the uncertainty must be far larger than the string scale, 
$\Delta x \sim v \gtrsim \sqrt{\alpha'}$, that is, 
$y \gtrsim 1$ where $m$ is larger than $1$. 
Since the other
non-extremal parameter $d$ is always smaller than 0 (see 
Fig.\ref{fig:d_Dinstanton.eps}), 
we conclude that the tachyon condensation
from unstable instantons to BPS saturated D$p$-branes is expressed by 
a three-parameter solution in the branch $I_{-}$: $m\ge 1,d\le 0$.

\subsubsection{Tachyonic/massive excitation on the D$p$\={D}$p$ system}

We have illustrated three types of deformation above. 
Here we would like to capture the general feature 
from above examples 
and discuss a possible generalization. 

First, we summarize the three cases. 
We considered the situation where the RR-charge is fixed as the same 
value as $N$ BPS D$p$-branes have.
In Case 1, the D$p$\={D}$p$ system with an tachyonic excitation 
gives the three-parameter solution with $d=0$ at low energy, 
which is trivial in some sense.
In Case 2, we considered the tachyon condensation 
on the D$9$\={D}$9$ system, which gives the solutions with $0 \le d$.
It is equivalent to the massive excitation on the BPS D$p$-branes.
On the other hand, in Case 3, although the tachyon condensation on the 
unstable D-instanton system is also equivalent to the massive excitation 
on the BPS D$p$-branes, their low energy solutions have $d \le 0$. 
The difference between Case 2 and Case 3 is the polarization of the massive 
vertex operators: 
The bosonic part of the vertex operator 
$\int d\widehat \sigma \bP_i D \bP^i$ 
is 
$\delta_{ij} \int d\sigma \del_\tau X^i \del_\tau X^j$, 
which describes the lowest massive mode of open strings, 
a rank $2$ tensor with zero momentum.
On the other hand, the vertex operator 
$\int d\widehat \sigma D\bX^\mu D^2\bX_\mu 
=\eta_{\mu\nu}\int d\sigma \del_\sigma X^\mu \del_\sigma X^\nu + \cdots$ 
has a longitudinal polarization. 
Physically, the effects of transverse (longitudinal) 
polarization is characterized by the fuzziness of the worldvolume 
in the transverse (longitudinal) direction, respectively.
In the context of the scattering of massive open strings, 
their difference is argued for example in \cite{Balasubramanian:1996xf}.
We here observe that the difference in the polarization 
is seen as the sign of $d$ if we probe these systems by 
massless closed strings.

Next, we discuss the possible generalization of the process of the 
tachyon condensation. 
In Case 2 and 3, we have only considered the tachyon condensation to BPS
D$p$-branes. 
However, we can also construct a D$p$\={D}$p$ system 
as long as they have the same RR-charge proportional to $N$, 
Therefore, 
recalling that a tachyon excitation on a system of unstable D9-branes or
unstable D-instantons is regarded as a massive excitation on the
resulting D$p$-branes (see Fig.~\ref{tachyon-massive}), 
we can combine the three cases as the D$p$\={D}$p$
system with tachyonic and massive excitations.  
As we have seen above, the tachyonic excitation contributes to $m$ and 
the massive excitations contribute to both $m$ and $d$. 
Note that $1\le m$ in any case.

It is then quite natural to take into account for higher massive excitations.
Let us consider vertex operators quadratic in $X$, say, 
\begin{equation}
\int d\sigma \del^l_\tau X^i \del^l_\tau X_i,\quad {\rm or}\quad 
\int d\sigma \del^l_\sigma X^\mu \del^l_\sigma X_\mu, 
\end{equation}
with $l=1,2,\cdots$. 
It is easy to show that such an excitation 
gives rise to the deformation of the boundary state 
in the form (\ref{general boundary state}).
It follows that they also contribute to the three-parameter 
solution
and gives another trajectory in the $(m,d)$ space.
Note that the relation between the polarization and the sign of $d$ 
is unchanged.
We can also consider vertex operators with higher power in $X$,
unless they break the global symmetry. 
In this case, the resulting deformed state no longer has the form 
(\ref{general boundary state})
but treating these vertex operators as perturbation, 
they contribute to the 
coupling to massless modes of closed strings. 
In any case, our conclusion is that the D$p$\={D}$p$ system with 
tachyonic and massive excitations are seen as the three-parameter 
solution at the low energy.

%%%%%%%%%%%%%%%%%%%%%%%%%%%%%%%%%
\section{Conclusions and Discussions}

In this paper, we discussed the stringy origin of the general 
solution of Type II supergravity with the symmetry $ISO(1,p)\times
SO(9-p)$, which is called as the three-parameter solution. 
This solution contains the BPS
saturated black $p$-brane solution in the parameter space, 
whose source is BPS saturated D$p$-branes expressed by a boundary state. 
We characterized the three-parameter solution in terms of 
two non-extremality parameter $m$ and $d$ on the supergravity side. 
On the other hand,
we discussed a class of the deformation of the boundary state 
on the string side. 
Then we determined the relation between the (macroscopic) 
non-extremality parameters of the classical
solution and the (microscopic) deformation parameters 
by extending the correspondence between the BPS black $p$-brane solution
and the boundary state. 
In particular, we showed that the dilaton charge is related to the
deformation of the boundary condition. 

We gave three examples of deformed boundary states
by considering the tachyon condensation. 
The first example is a D$p$\=D$p$-brane system with a constant tachyon VEV 
discussed in \cite{Kobayashi:2004ay}, 
the second and the third example are the tachyon condensation 
processes from the unstable D9-branes and the unstable D-instantons 
to the BPS D$p$-branes, respectively. 
In the latter two examples, the boundary condition in the
longitudinal and the transverse directions are deformed, respectively, 
then the corresponding classical solution learns to possess
a non-trivial dilaton charge. 
We also showed that the deformed systems are generally regarded as 
tachyonic and/or massive excitations of the open strings
on D$p$\={D}$p$-brane systems.

Our method is also applicable to the charge-neutral case and/or 
more complicated (less symmetric) systems. 
For example, the intersecting D-brane system is the one with several 
RR-charges and less global symmetry \cite{Miao:2004bn,Rama:2005bd}. 
Another possible application is 
the study of the relationship between 
the stability of the supergravity solution and that
of the D-brane system. 
In this paper, we only consider a static solution, thus the source is
also static even if tachyon fields are excited. 
However, if we consider a perturbation from the solution, 
the unstable modes are expected to correspond to tachyonic excitations on the D-branes. 
This would give the stringy meaning of the instability of the 
supergravity solutions.

It is also interesting to investigate the properties of 
the dilaton charge further. 
As we repeated in this paper, 
the existence of the dilaton charge changes the structures of spacetimes, 
in particular, it generally makes the spacetimes have no horizons. 
Therefore,  
the study of the dilaton charge from the viewpoint of string theory
might lead us to the understanding of the meaning of the horizons from 
the viewpoint of the string theory. 
However, since the three-parameter 
solution does not have the horizon in most region of the parameter
space, 
we will have to treat the four-parameter solution 
which has the horizon in some parameter region  
\cite{Zhou:1999nm, Brax:2000cf}, 
in order to play with the most interesting feature of black objects. 
For this purpose, our strategy is expected to be essentially applicable. 
If we can clarify this issue, 
it might be possible, 
for example, to understand the Schwarzschild black hole in 
terms of the string theory, which would become one of the points of contact 
for the string theory and the general relativity.

%%%%%%%%%%%%%%%%%%%%%%%%%%%%%%%%%%%%%%%%%%%%%%%%%%%%%%%%%%%%
\section*{Acknowledgments}
The authors would like to thank 
T.~Harada, 
M.~Hayakawa, 
Y.~Himemoto, 
D.~Ida,
Y.~Ishimoto,  
G.~Kang, 
H.~Kawai, 
S.~Kinoshita, 
H.~Kudoh, 
Y.~Kurita,
K.~Maeda,
Sh.~Matsuura, 
S.~Mukohyama,
K.~Nakao, 
M.~Natsuume,  
K.~Ohta, 
N.~Ohta, 
T.~Onogi, 
M.~Sakagami, 
N.~Sasakura,
J.~Soda,  
K.~Takahashi,
T.~Tada, 
T.~Tamaki,
S.~Watamura, 
B.~de Wit 
and 
J.~Yokoyama 
for great supports and helpful discussions. 
This work of T.A. and S.M. is supported 
by Special Postdoctoral Researchers
Program at RIKEN.

%%%%%%%%%%%%%%%%%%%%%%%%%%%%%%%%%%%%%%%%%%%%%%%%%%%%%%%%%%%%

\appendix

\section{Construction of Boundary States}
\label{sec:costruction}

\subsection{Construction from D9-branes}
In this appendix, we review the tachyon condensation from 
a D9\=D9 system to $N$ BPS D$p$-brane  
in Type IIB superstring theory \cite{Asakawa:2002ui}. 

We start with a off-shell boundary state corresponding to 
D9\=D9-brane system in the
NSNS-sector on which the tachyon field is excited;%
\begin{equation}
\ket{Bp'}_{\rm NS} \equiv  e^{-S_b(T)}\ket{B9}_{\rm NS}, 
\label{gaussian boundary state}
\end{equation}
where $\ket{B9}_{\rm NS}$ is the boundary state of a single D9-brane in the
NSNS-sector (\ref{BPS boundary state}) and 
\begin{equation}
e^{-S_b}={\Tr}\, \widehat{\rm P} e^{\int d\hsigma \bM(\hsigma)}, \qquad
\bM=\left(
\begin{array}{cc}
0 & T(\bX) \\ 
T^{\dagger}(\bX) & 0
\end{array}
\right), 
\end{equation}
is a boundary interaction 
and $\widehat{\rm P}$ denotes the supersymmetric path ordered
product. 
%\footnote{
$\bX^M (\hsigma)$ and $\bP^M (\hsigma)$ denote the position boundary
superfields and the conjugate momentum superfields on the boundary, 
respectively, 
and $\hsigma=(\sigma,\theta)$ is the boundary supercoordinate. 
For notation of the superfields and the supercoordinate, see
\cite{Friedan:1985ge}.  
For construction of the boundary state, see \cite{Callan:1987px}. 
%}
When $\bM$ can be expanded by $SO(2m)$ gamma matrices $\Gamma^I$ 
$(I=1,\cdots,2m)$ as 
\begin{equation}
\bM=\sum_{k=0}^{2m}\bM^{I_1\cdots I_{k}}\otimes \Gamma^{I_1\cdots I_k}, 
\end{equation}
it is convenient to rewrite 
the boundary interaction using fermionic superfields as \cite{Kraus:2000nj}% 
\footnote{
The $\Tr{\rm P}$ in the second line is necessary when 
$\bM_{I_1\cdots I_k}$ are also matrices. 
} 
\begin{align}
 e^{-S_b}
 &= \int \left[d\bGamma^I\right]
 \Tr {\rm P}
 \exp\left\{
 \int d\hsigma\left(
  \frac{1}{4}\bGamma^I D \bGamma^I 
  +\sum_{k=0}^{2m}\bM_{I_1\cdots I_k}
   \bGamma^{I_1}\cdots\bGamma^{I_k}
 \right)
 \right\}. 
\label{gamma decomposition}
\end{align}
We fix the measure of the path integral so that the boundary interaction
(\ref{gamma decomposition}) gives the number of D9-branes 
in the absence of the tachyon field.

Suppose $N 2^{(9-p)/2}$-pairs of D9-brane and \=D9-brane 
and consider the tachyon profile, 
\begin{equation}
\bM = u \Gamma^i \bX^i\,\otimes {1}_{N}, 
\end{equation}
where $\Gamma^i$ $(i=p+1,\cdots,9)$ are the $SO(2(9-p))$
$\gamma$-matrices. 
Then (\ref{gaussian boundary state}) becomes 
\begin{align}
\ket{Bp;u}_{\rm NS} &= \int [d\bX^{M}d\bGamma^i] 
\Tr P \exp\left[\int d\hsigma \left(
\frac{1}{4}\bGamma^i D \bGamma^i + u \bX^i\bGamma^i
\right)\right]\ket{\bX^M}_{\rm NS},  
\end{align}
where the measure $[d\bX^M]$ is determined so that this state becomes 
the boundary state of $N 2^{(9-p)/2}$-pairs of D9-brane and \=D9-brane 
in the limit of $u\to0$. 
Since $\ket{\bX} = e^{\int d\hsigma i\bX \bP} \ket{\bX=0}$ using the
conjugate momentum superfield $\bP$, 
the boundary fermion fields $\bGamma^i$ are replaced by 
the momentum superfields $\bP^i$ by carrying out the functional integral
for $\bX^i$ and $\bGamma^i$. Moreover, it is easy to check that this
state imposes the Neumann boundary condition for the directions 
$\mu=0,\cdots,p$. Then (\ref{gaussian boundary state}) can be written
as   
\begin{align}
\ket{Bp';u}_{\rm NS} &= N \exp\left[
\int d\hsigma \left(
-\frac{1}{4u^2}\bP_i D\bP_i
\right)
\right]\ket{Bp}_{\rm NS},
\label{gaussian BS} 
\end{align}
where we determine the constant factor of (\ref{gaussian boundary
state}) so that $\ket{Bp';u}_{\rm NS}$ becomes the boundary state corresponding
to the NSNS-sector 
of $N$ D$p$-branes in the limit of $u\to\infty$.

Using the explicit expression of $\bP_i(\hsigma)$ by the string oscillators and 
the zeta-function regularization, 
\begin{align}
\prod_{m=1}^\infty\frac{b}{m+a} = \frac{\Gamma(a+1)}{\sqrt{2\pi b}}, 
\qquad 
\prod_{r=1/2}^\infty\frac{b}{r+a} = \frac{\Gamma(a+1/2)}{\sqrt{2\pi}}, 
\end{align}
we obtain 
\begin{align}
 \ket{Bp';u}_{\rm NS} = 
 &N \frac{T_p}{2}
 \left[\frac{4^y y^{1/2}\Gamma^2(y)}{\sqrt{4\pi}\Gamma(2y)}\right]^{9-p}
 \exp\left[{-\sum_{n=1}^{\infty} 
 \frac{1}{n} \alpha_{-n}^M S^{(n)}_{MN} \talpha_{-n}^N} 
 +i\sum_{r=1/2}^{\infty} b_{-r}^M S^{(r)}_{MN} \tb_{-r}^N 
 \right] \times \nn 
  &\times \int \frac{\prod_{i=p+1}^{9} dp_i}{(2\pi)^{9-p}} 
  \exp\left(-\frac{p_i^2}{8\pi u^2}\right)
 \ket{p_\mu=0,\,p_i}_{\rm NS}, \qquad (y=4\pi\alpha'u^2)
 \label{gaussian BS 2}
\end{align}
with
\begin{equation}
 S^{(n)}_{MN} 
     = \left(\eta_{\mu\nu},\, -\frac{y-n}{y+n} \delta_{ij} \right), 
 \quad 
 S^{(r)}_{MN} = \left(\eta_{\mu\nu},\, -\frac{y-r}{y+r}
		     \delta_{ij} \right). 
\end{equation} 
We can easily show that 
(\ref{gaussian BS 2}) correctly becomes 
the boundary state of the NSNS-sector of the $N$ D$p$-branes 
in the limit of $u\to\infty$, 
which is apparent from the original definition 
(\ref{gaussian BS}). 
On the other hand, in the limit of $u\to0$, (\ref{gaussian BS 2})
becomes 
\begin{equation}
\ket{Bp';u\to0}= 2^{(9-p)/2}N \frac{(4\pi^2\alpha')^{-(9-p)/2}T_p}{2}
e^{
-\sum_{n=1}^{\infty} 
 \frac{1}{n} \alpha_{-n}^M \talpha_{-n}^M
 +i\sum_{r=1/2}^{\infty} b_{-r}^M \tb_{-r}^M
}\ket{p_M=0}_{\rm NS}. 
\label{gaussian BS u=0}
\end{equation}
Since $(4\pi^2\alpha')^{-(9-p)/2}T_p=T_9$, (\ref{gaussian BS 2}) correctly 
expresses the tachyon
condensation from $N 2^{(9-p)/2}$-pairs of  
D9-brane and \=D9-brane to $N$ D$p$-branes. 

For the RR sector, we can carry out the similar calculation 
and a deformed boundary state like (\ref{gaussian BS 2}) appears, 
that is, the boundary condition and the zero-mode part is deformed.
However, in the case of the RR-sector, the normalization factor of the
state does not depend on $u$ since the contribution from the bosonic
oscillators and the fermionic oscillators completely cancel. 
This leads to the conservation of the RR charge 
under the tachyon condensation.

\subsection{Construction from D-instantons}
Next, we review tachyon condensation of infinitely many 
non-BPS D-instantons to the $N$ D$p$-branes. 
For detail, see \cite{Asakawa:2002ui}. 

In Type IIA superstring theory, a state corresponding to non-BPS
D-instantons with scalar profiles and a tachyon profile is given by 
\begin{gather}
 \ket{Bp'}=e^{-S_b(\Phi^\mu,T)}\ket{B(-1)}_{\rm NS}, 
\end{gather}
where 
\begin{align}
 e^{-S_b} = {\rm Tr}\widehat{\rm P}e^{
  \int d\widehat{\sigma} {\mathbf M}(\widehat{\sigma})
 }, \qquad
 \bM = \left(
 \begin{array}{cc}
  -i\Phi^M\bP_M & T \\
  T & -i\Phi^M\bP_M 
 \end{array}
 \right), 
\end{align}
is a boundary interaction.

The solution that corresponds to $N$ D$p$-branes is%
\begin{align}
 T = v \hp_\mu\otimes 1_N \otimes \gamma^\mu, \qquad 
 \Phi^\mu = \hx^\mu \otimes 1_N \otimes 1, \qquad 
 \Phi^i =0, 
\label{Dp config}
\end{align}
where $\gamma^\mu$ are gamma matrices, and 
$\hx^\mu$ and $\hp_\mu$ are operators satisfying 
\begin{equation}
 \left[\hx^\mu,\hp_\nu\right]
 = i\delta^\mu_\nu.
\end{equation}
In the configuration (\ref{Dp config}), the matrix $\bM$ 
is decomposed as 
\begin{equation}
 \bM = \left(-i\hx^\mu\bP_\mu\right) {1} 
      +\left(v\hp_\mu\right) \Gamma^\mu, 
 \qquad 
  \Gamma^\mu = \left(
  \begin{array}{cc}
   0 & \gamma^\mu \\
  \gamma^\mu & 0
  \end{array}
  \right)
\end{equation}
then it is again convenient to use the boundary fermion as 
(\ref{gamma decomposition}); 
\begin{align}
 e^{-S_b} &=  
 \int \left[d\bGamma^\mu\right]
 \Tr \widehat{\rm P}
 \exp\left\{
 \int d\hsigma\left(
  \frac{1}{4}\bGamma^\mu D \bGamma^\mu 
  -i\hx^\mu\bP_\mu 
      +\left(v\hp_\mu\right) \bGamma^\mu
 \right)\right\} \nn
 &= N\int \left[d\bGamma^\mu
       d \bx^\mu d \bp_\mu \right]
 \exp\left\{
 \int d\hsigma\left(
  \frac{1}{4}\bGamma^\mu D \bGamma^\mu 
  +i \bp_\mu D\bx^\mu 
  -i \bx^\mu\bP_\mu
      + v\bp_\mu \bGamma^\mu
 \right)\right\} \nn
 &= N\int \left[d\bGamma^\mu d \bx^\mu \right]
 \delta\left(D\bx^\mu -i v \bGamma^\mu \right)
 \exp\left\{
 \int d\hsigma \left(
 \frac{1}{4}\bGamma^\mu D \bGamma^\mu 
 -i\bx^\mu \bP_\mu
 \right)\right\} \nn 
 &= N\int \left[d\bX^\mu \right]
 \exp\left\{
 \int d\hsigma \left(
  -i\bX^\mu\bP_\mu 
  -\frac{1}{4v^2}D\bX^\mu D^2\bX^\mu
 \right)\right\}. 
\end{align}
Here we have replaced the operators $\hx^\mu$ and 
$\hp_\mu$ by superfields $\bx^\mu(\hsigma)$ and 
$\bp_\mu(\hsigma)$ in the second line 
by adding the kinetic term $i\bp_\mu D\bx^\mu$. 
Then we performed the functional integral for 
$\bp_\mu$ and $\bGamma^\mu$.
In the last line $\bx^\mu$ is identified with the superfield $\bX^\mu$ 
on the string worldsheet. 
Then the state corresponding to this solution is 
\begin{align}
 \ket{Bp';v}_{\rm NS} &\equiv  
 \int \left[d\bX^\mu \right]
 \exp\left\{
 \int d\hsigma \left(
  -i\bX^\mu\bP_\mu 
  -\frac{1}{4v^2}D\bX^\mu D^2\bX^\mu
 \right)\right\}\ket{\bX^M=0}_{\rm NS} \nn
 &= 
 \int \left[d\bX^\mu \right] 
  \exp\left\{
 \int d\hsigma \left(
  -\frac{1}{4v^2}D\bX^\mu D^2\bX^\mu
 \right)\right\}\ket{\bX^\mu,\bX^i=0}_{\rm NS}, 
\end{align}
where the measure has been fixed so that
this state expresses the boundary state of $N$ D$p$-branes 
in the $v\to\infty$ limit.  
This expression can again be evaluated using 
the zeta-function regularization as 
\begin{align}
\ket{Bp';v}_{\rm NS} &= 
\frac{NT_p}{2}
\left(\frac{4^y y^{1/2}\Gamma^2(y)}{\sqrt{4\pi}\Gamma(2y)}\right)^{p+1} \nn
&\hspace{5mm}\times
\exp\left[
-\sum_{n=1}^\infty\frac{1}{n}
\alpha^M_{-n} S^{(n)}_{MN}\talpha^N_{-n}
+i\sum_{r=1/2}^\infty b_{-r}^M S_{MN}^{(r)} \tb_{-r}^N
\right]\ket{p_\mu=0,x^i=0}_{\rm NS},
\end{align}
where $y=v^2/\pi \alpha'$ and 
\begin{equation}
 S_{MN}^{(n)} = 
\left(\frac{y-n}{y+n}\eta_{\mu\nu},-\delta_{ij}\right), 
\qquad 
 S_{MN}^{(r)} = 
\left(\frac{y-r}{y+r}\eta_{\mu\nu},-\delta_{ij}\right), 
\end{equation} 
In the limit of $y\to\infty$, 
\begin{equation}
\frac{4^y y^{1/2}\Gamma^2(y)}{\sqrt{4\pi} \Gamma(2y)} \to 1 
\end{equation}
as shown in Fig.\ref{fig:F.eps}.

\bibliographystyle{JHEP}
\bibliography{refs}

\end{document}